\documentclass[12pt]{article}
\usepackage{eurosym}
\usepackage{graphicx}
\usepackage[utf8]{inputenc}
\usepackage[T1]{fontenc}
\usepackage{indentfirst}
\usepackage[margin=0.6in,nomarginpar]{geometry}
\usepackage[final]{hyperref}
\usepackage{amsmath}
\usepackage[overload]{empheq}
\usepackage{hyperref}
\usepackage{cite}
\usepackage{xcolor}
\usepackage{subcaption}
\usepackage{caption}
\usepackage{amssymb}
\usepackage{multirow}
\usepackage[table]{xcolor}
\usepackage{orcidlink}
\usepackage{float}
\usepackage{hyperref}
\usepackage{graphicx}

\hypersetup{
colorlinks=true,
linkcolor=blue,
citecolor=blue,
filecolor=magenta,
urlcolor=blue
}

\begin{document}

\title{Time-Dependent Dunkl-Pauli Oscillator in the Presence of the
Aharonov-Bohm Effect}
\author{ Boubakeur Khantoul\,\orcidlink{0000-0001-9012-4864}$^{1,2}$\thanks{%
Email: \texttt{boubakeur.khantoul@univ-constantine3.dz}} \and Ahmed Tedjani\,%
\orcidlink{0000-0002-9590-4082}$^{1}$\thanks{%
Email: \texttt{ahmed.tedjani@univ-constantine3.dz}} \\
$^{1}$Department of Process Engineering, University of Constantine 3 -Salah
Boubnider, \\
25016 Constantine, Algeria\\
$^{2}$Theoretical Physics Laboratory, Department of Physics, University of
Jijel, Algeria }
\maketitle

\begin{abstract}
We present an exact, time-dependent solution for a two-dimensional Pauli
oscillator deformed by Dunkl operators in the presence of an Aharonov--Bohm
(AB) flux. By replacing conventional momenta with Dunkl momenta and allowing
arbitrary time dependence in both, mass and frequency, we derive a deformed
Pauli Hamiltonian that encodes reflection symmetries and topological gauge
phases. Employing the Lewis-Riesenfeld invariant method, we derive exact
expressions for the eigenvalues and spinor eigenfunctions of the system.
Crucially, the AB flux imposes symmetry constraints on the Dunkl parameters
of the form $\nu_1 = \mp \nu_2 $, linking the reflection symmetry ($\epsilon
= \pm 1 $) to the quantization of angular momentum. These constraints modify
the energy spectrum and wavefunctions of the angular operator and the
invariant operator. Our framework reveals novel spectral characteristics
arising from the interplay between topology and Dunkl symmetry, with
potential implications for quantum simulation in engineered systems such as
cold atoms and quantum dots.
\end{abstract}



\vspace{1cm}

\section*{Introduction}

The quantum mechanical description of spin-1/2 particles interacting with
electromagnetic fields is elegantly captured by the Pauli equation \cite{1}.
This fundamental equation, a non-relativistic limit of the Dirac equation 
\cite{2,3}, incorporates the intrinsic magnetic moment of particles and
their interaction with external magnetic fields, leading to crucial
phenomena such as spin precession \cite{4} and the Zeeman effect \cite{5}.
The Pauli equation forms the bedrock for understanding a vast array of
physical phenomena, including the behavior of electrons in atoms and
molecules \cite{6,7}, the principles underlying nuclear magnetic resonance
(NMR) \cite{8,9} and electron spin resonance (ESR) spectroscopy \cite{10,11}%
, and the transport properties of electrons in materials \cite{12,13}. For
instance, the fine structure of atomic spectra, arising from spin-orbit
coupling \cite{14,15}, is directly explained by terms within the Pauli
Hamiltonian. Moreover, the Pauli equation is essential in understanding the
quantum Hall effect \cite{16,17} and the magnetic properties of condensed
matter systems, including topological insulators and superconductors \cite%
{18,19}. Recent research continues to explore relativistic corrections to
the Pauli equation and its applications in high-precision measurements \cite%
{21}. Furthermore, various symmetry approaches have been used to investigate
the Pauli equation, including supersymmetric factorization techniques \cite%
{16A,17A,18A,19A} and the analysis of Lie and discrete symmetries \cite%
{20A,21A}.

While the standard formulation of quantum mechanics relies on the standard
derivatives, the exploration of alternative differential operators has
unveiled new perspectives on fundamental quantum descriptions \cite{22}.
Among these, Dunkl derivatives stand out due to their intimate connection
with reflection groups and their ability to incorporate symmetries beyond
simple translations \cite{23,24,37A}. These operators, which generalize
standard derivatives by including reflection terms, naturally arise in the
study of integrable systems \cite{25,26,33A}, quantum many-body problems
with exchange interactions \cite{27,28}, and quantum mechanics in spaces
with specific symmetries, such as root systems \cite{29}. The use of Dunkl
derivatives allows for the construction of alternative quantum mechanical
equations that can describe systems with non-trivial exchange statistics
(e.g., anyons) \cite{31,32} or in spaces where reflection symmetries play a
significant role, such as in confined geometries or with specific boundary
conditions \cite{33}. For example, Dunkl oscillators, whose dynamics is
governed by Hamiltonians that involve Dunkl derivatives, exhibit energy
spectra and wave functions that differ significantly from their standard
counterparts, providing insights into the role of exchange interactions and
generalized statistics \cite{35,36,61A,62A}. Furthermore, Dunkl operators
have found applications in areas such as mathematical physics,
representation theory, and the study of special functions, highlighting
their broad significance in extending our understanding of mathematical and
physical structures \cite{40A,41A,42A}. Recent advances include the study of
Dunkl-type equations in higher dimensions and their connection to fractional
quantum mechanics \cite{39,40,44}.

In the present study, we extend this framework by formulating and analyzing
the time-dependent Pauli equation in the context of a harmonic oscillator,
further enriched by the inclusion of Aharonov-Bohm (AB) effects and
employing Dunkl derivatives \cite{47,48,50}. The harmonic oscillator, a
cornerstone model in quantum mechanics with applications ranging from
molecular vibrations \cite{51} to quantum field theory \cite{52}, provides a
well-defined system to explore the impact of these generalizations. Recent
studies have investigated the behavior of quantum harmonic oscillators in
non-inertial frames and under the influence of external fields \cite{54}.
The AB effect, a quintessential quantum phenomenon demonstrating the
non-local influence of electromagnetic potentials on charged particles even
in regions where the magnetic field is zero \cite{55,56,23A,47A}, adds
another layer of complexity and richness to the system, revealing
fundamental aspects of gauge invariance and quantum connectivity \cite{57,58}%
. Recent research has explored the AB effect in mesoscopic systems and
topological materials \cite{59,60}. By substituting the standard derivatives
with Dunkl derivatives in the Pauli equation, we aim to investigate how
these generalized operators, reflecting specific underlying symmetries,
alter the spin dynamics and spatial behavior of the charged harmonic
oscillator in the presence of an AB field \cite{61,62}. This approach
connects the intrinsic spin of the particle with the non-trivial spatial
symmetries encoded in the Dunkl operators and the topological aspects
introduced by the AB effect.

To address this complex problem, we employ the invariant method, a powerful
technique for solving time-dependent quantum systems\cite{64}. More
recently, it has been applied in the context of Dunkl-type operators in
time-dependent quantum models \cite{64BC,64D}. This approach enables the
identification of conserved quantities (invariants), which simplify the
resolution of the Schr\"{o}dinger-like equation. Recent developments have
extended the invariant method to a wide range of time-dependent Hamiltonians
and open quantum systems. Subsequently, we analyze the eigenvalue equation
for the invariant, paying particular attention to the connection points
arising from the AB effect \cite{67,67A,67B}. Understanding the behavior of
wave functions around these points, where the vector potential might exhibit
singularities or discontinuities, is crucial to grasping the physical
implications of the combined effects of spin, Dunkl derivatives, and
topological phases \cite{68,69}. Finally, we derive the phase (Berry phase
and dynamic phase) \cite{70,71} and construct the general solution for the
time-dependent Dunkl-Pauli equation for the harmonic oscillator under AB
influence. This study offers novel insights into the behavior of quantum
systems in non-standard settings and has the potential to broaden our
understanding of fundamental quantum phenomena, including the interplay of
spin, symmetry, and topology in quantum mechanics \cite{72,73}. Further
research in this direction could explore applications in quantum information
processing and the design of novel quantum devices \cite{74,75}.

\section*{2D TD Dunkl-Pauli oscillator in presence of AB effect}

We consider the two-dimensional, time-dependent Pauli equation describing a
spin-$\frac{1}{2}$ particle subjected to both a harmonic oscillator
potential and the Aharonov--Bohm (AB) effect. This formulation arises
naturally as the non-relativistic limit of the Dirac equation in the
presence of electromagnetic interactions. The governing equation takes the
form

\begin{equation}
\left[ \frac{1}{2M\left( t\right) }\left( \overrightarrow{\sigma _{j}}.%
\overrightarrow{\pi _{j}}\right) ^{2}-\frac{eB(r)}{2M\left( t\right) }\sigma
_{z}+\frac{1}{2}M\left( t\right) \omega \left( t\right) ^{2}\left(
x^{2}+y^{2}\right) \right] \psi \left( x,y,t\right) =i\frac{\partial }{%
\partial t}\psi \left( x,y,t\right)  \label{1}
\end{equation}%
where $\sigma _{j}$ are the Pauli matrices and $\overrightarrow{\pi }$
denotes the gauge-invariant momentum operator, given by 
\begin{equation}
\overrightarrow{\pi }=\overrightarrow{p}-e\overrightarrow{\mathbf{A}}=\left(
p_{x}-eA_{x},p_{y}-eA_{y}\right)  \label{2}
\end{equation}%
with $e$ being the electric charge and the speed of light set to $c=1.$The
two-component spinor wave function is expressed as%
\begin{equation}
\psi =\binom{\psi _{1}}{\psi _{2}}  \label{3}
\end{equation}

To account for the Aharonov-Bohm (AB) effect, we consider a magnetic field $%
\mathbf{B}$ confined to an infinitely thin solenoid aligned along the $z$%
-axis, perpendicular to the plane of particle motion. The field is singular
and localized at the origin, with its spatial distribution described by 
\begin{equation}
e\mathbf{B}(r)=\frac{\vartheta }{r}\delta (r),
\end{equation}%
where $\vartheta $ denotes the total magnetic flux threading the filament,
assumed to be finite and non-zero. In the Coulomb gauge, the corresponding
vector potential $\mathbf{A}$ associated with this configuration, for a flux
tube of zero radius (infinitely small), takes the form 
\begin{equation}
e\mathbf{A}=-\frac{\vartheta }{r}\overrightarrow{u}_{\varphi },
\end{equation}%
where $\overrightarrow{u}_{\varphi }$ is the azimuthal unit vector in polar
coordinates.

The vector potential can be rewritten explicitly in Cartesian coordinates as 
\cite{75A,75B,75C}

\begin{equation}
e\mathbf{A}=-\frac{\vartheta }{r}\overrightarrow{u}_{\varphi }=-\frac{%
\vartheta }{r}\left( -\sin \varphi \overrightarrow{i}+\cos \varphi 
\overrightarrow{j}\right) =\frac{\vartheta y}{\left( x^{2}+y^{2}\right) }%
\overrightarrow{i}-\frac{\vartheta x}{\left( x^{2}+y^{2}\right) }%
\overrightarrow{j}.  \label{3-b}
\end{equation}

To construct the Dunkl-Pauli equation (DPE), the standard momentum operator
is replaced by the Dunkl momentum operator, defined as 
\begin{equation}
p_{j}=\frac{1}{i}D_{j},  \label{4}
\end{equation}%
where $D_{j}$ denotes the Dunkl derivative along the direction $x_{j}$,
given by%
\begin{equation}
D_{j}=\frac{\partial }{\partial x_{j}}+\frac{\nu _{j}}{x_{j}}\left(
1-R_{j}\right) ,  \label{5}
\end{equation}%
here, $\nu _{j}$ are real deformation (or Wigner) parameters satisfying the
condition $\nu _{j}>-\frac{1}{2}$, \ and $R_{j}$ are reflection operators
acting on a function $f(x)$ as

\begin{equation}
R_{j}f\left( x\right) =f\left( ...,-x_{j},...\right) ;\text{ \ \ }%
R_{j}x_{i}=-\delta _{ij}x_{j}R_{j};\text{ \ \ }R_{i}R_{j}=R_{j}R_{i}\text{
and }R_{j}^{2}=1.  \label{7}
\end{equation}%
\ \ \ 

The square of the Dunkl derivative yields a second-order
differential-difference operator of the form%
\begin{equation}
D_{j}^{2}=\frac{\partial ^{2}}{\partial x_{j}^{2}}+\frac{2\nu _{j}}{x_{j}}%
\frac{\partial }{\partial x_{j}}-\frac{\nu _{j}}{x_{j}^{2}}\left(
1-R_{j}\right) .  \label{6}
\end{equation}%
These operators generate a deformed Heisenberg algebra, characterized by the
following commutation relations 
\begin{equation}
\left[ x_{i},D_{j}\right] =\delta _{ij}\left( 1+2\nu _{j}R_{j}\right) ;\text{
\ \ }\left[ D_{i},D_{j}\right] =\left[ x_{i},x_{j}\right] =0,  \label{8}
\end{equation}

Incorporating these structures, the Dunkl-Pauli Hamiltonian takes the form%
\begin{multline}
H = -\frac{1}{2M}\triangle _{D} +\frac{1}{2M}\frac{\vartheta ^{2}}{%
x^{2}+y^{2}} +\frac{1}{2M}\frac{1}{i} \Bigg[ \frac{2\vartheta x}{x^{2}+y^{2}}
\frac{\partial }{\partial y} + \frac{2\vartheta x}{x^{2}+y^{2}} \frac{\nu
_{2}}{y}(1-R_{2}) - \frac{2\vartheta y}{x^{2}+y^{2}} \frac{\partial }{%
\partial x} \\
- \frac{2\vartheta y}{x^{2}+y^{2}} \frac{\nu _{1}}{x}(1-R_{1}) \Bigg] + 
\frac{1}{2M} \sigma _{z} \left[ \frac{2\vartheta}{x^{2}+y^{2}}(\nu
_{1}R_{1}+\nu _{2}R_{2}) \right] - \frac{\vartheta }{2M} \frac{\delta(r)}{r}
\sigma _{z} + \frac{1}{2} M \omega ^{2}(x^{2}+y^{2})
\end{multline}

where the Dunkl Laplacian is defined as:%
\begin{equation}
\triangle _{D}=\frac{\partial ^{2}}{\partial x^{2}}+\frac{\partial ^{2}}{%
\partial y^{2}}+\frac{2\nu _{1}}{x}\frac{\partial }{\partial x}+\frac{2\nu
_{2}}{y}\frac{\partial }{\partial y}-\frac{\nu _{1}}{x^{2}}\left(
1-R_{1}\right) -\frac{\nu _{2}}{y^{2}}\left( 1-R_{2}\right)  \label{12}
\end{equation}

In polar coordinates, where $x=r\cos \varphi ,$ and $y=r\sin \varphi ,$ the
Hamiltonian can be rewritten as:%
\begin{align}
H\left( t\right) & =-\frac{1}{2M\left( t\right) }\frac{\partial ^{2}}{%
\partial r^{2}}-\frac{1+2\nu _{1}+2\nu _{2}}{2M\left( t\right) r}\frac{%
\partial }{\partial r}+\frac{\mathcal{B}_{\varphi }}{M\left( t\right) r^{2}}+%
\frac{1}{2}M\omega ^{2}r^{2}  \notag \\
& -\frac{\vartheta }{Mr^{2}}\mathcal{J}_{\varphi }+\left[ \frac{\vartheta }{%
Mr^{2}}\left( \nu _{1}R_{1}+\nu _{2}R_{2}\right) -\frac{\vartheta }{2M}\frac{%
\delta \left( r\right) }{r}\right] \sigma _{z}  \label{14}
\end{align}

The operators $\mathcal{B}_{\theta }$ and $\mathcal{J}_{\theta },$
representing the Dunkl angular operators, are defined as: \cite{76,77,78,79}:%
\begin{equation}
\mathcal{B}_{\varphi }=-\frac{1}{2}\frac{\partial ^{2}}{\partial \varphi ^{2}%
}+\left( \nu _{1}\tan \varphi -\nu _{2}\cot \varphi \right) \frac{\partial }{%
\partial \varphi }+\frac{\nu _{1}}{2\cos ^{2}\varphi }\left( 1-R_{1}\right) +%
\frac{\nu _{2}}{2\sin ^{2}\varphi }\left( 1-R_{2}\right)  \label{16}
\end{equation}%
\begin{equation}
\mathcal{J}_{\varphi }=i\left( \frac{\partial }{\partial \varphi }+\left[
\nu _{2}\cot \varphi \left( 1-R_{2}\right) -\nu _{1}\tan \varphi \left(
1-R_{1}\right) \right] \right)
\end{equation}%
Using the following relation 
\begin{equation}
\mathcal{J}_{\varphi }^{2}=2\mathcal{B}_{\varphi }+2\nu _{1}\nu _{2}\left(
1-R_{1}R_{2}\right)  \label{20}
\end{equation}%
and the expressions for the radial momentum operators:%
\begin{eqnarray}
p_{r} &=&-i\left( \frac{\partial }{\partial r}+\frac{\delta }{r}\right) , \\
p_{r}^{2} &=&-\frac{\partial ^{2}}{\partial r^{2}}-\frac{2\delta }{r}\frac{%
\partial }{\partial r}-\frac{\delta \left( \delta -1\right) }{r^{2}}
\end{eqnarray}%
where%
\begin{equation}
\delta =\frac{1}{2}+\nu _{1}+\nu _{2},
\end{equation}%
then, the Hamiltonian can be rewritten as%
\begin{equation}
\resizebox{0.95\textwidth}{!}{$ H(t)=\frac{1}{2M}\left(
p_{r}^{2}+\frac{\vartheta^{2}-2\vartheta
\mathcal{J}_{\varphi}+\mathcal{J}_{\varphi}^{2}+\delta(\delta-1)-2\nu_{1}%
\nu_{2}(1-R_{1}R_{2})+2\vartheta(\nu_{1}R_{1}+\nu_{2}R_{2})%
\sigma_{z}}{r^{2}}-\vartheta\frac{\delta(r)}{r}\sigma_{z}\right)+%
\frac{1}{2}M\omega^{2}r^{2} $}  \label{21}
\end{equation}
To obtain exact solutions of the Schr\"{o}dinger equation associated with
the Hamiltonian presented in $\left( \ref{21}\right) $, we adopt the
Lewis--Riesenfeld invariant method. This approach provides a systematic
framework for solving time-dependent quantum systems and is particularly
effective in addressing the time-dependent Pauli equation. Within this
formalism, the Hamiltonian and the corresponding invariant operator are
required to satisfy the Lewis--Riesenfeld invariant equation, ensuring the
consistency of the dynamical evolution:%
\begin{equation}
\frac{dI(t)}{dt}=\frac{\partial I(t)}{\partial t}+\frac{1}{i\hbar }%
[I(t),H(t)]=0.  \label{LW}
\end{equation}

The solution of the time-dependent Schr\"{o}dinger equation, denoted by $%
\psi (r,\theta ,t)$, can be constructed from the eigenfunctions of an
invariant operator $I\left( t\right) $. Specifically, the eigenvalue
equation takes the form%
\begin{equation}
I(t)\mathcal{F}(r,\varphi ,t)=E_{n,l,m_{s}}\mathcal{F}(r,\varphi ,t),
\label{EVE}
\end{equation}%
where $\mathcal{F}(r,\varphi ,t)$ is the eigenfunction associated with the
eigenvalue $E_{n,l,m}$ . The full time-dependent wave function is then
expressed as 
\begin{equation}
\psi (r,\varphi ,t)=e^{i\eta (t)}\mathcal{F}(r,\varphi ),
\end{equation}%
where $\eta (t)$ is the quantum phase, which can be determined from the
equation 
\begin{equation}
\hbar \frac{d}{dt}\eta (t)=\langle \mathcal{F}(r,\varphi )|i\hbar \frac{%
\partial }{\partial t}-H|\mathcal{F}(r,\varphi )\rangle .  \label{PH}
\end{equation}

To construct the exact Lewis--Riesenfeld invariant corresponding to the
system described in Eq. $\left( \ref{21}\right) $, we introduce a set of
generators $\left\{ T_{1},T_{2},T_{3}\right\} $, explicitly defined as: 
\begin{equation}
\begin{aligned} \left\{ \begin{aligned} T_{1} &= \frac{1}{2}p_{r}^2 +
\frac{\vartheta^2 - 2\vartheta J_{\phi} + J_{\phi}^2 + \delta(\delta - 1) -
2\nu_1\nu_2\bigl(1 - R_1R_2\bigr) + 2\vartheta\bigl(\nu_1R_1 +
\nu_2R_2\bigr)\sigma_{z}}{r^2} - \vartheta\,\frac{\delta(r)}{r}\,\sigma_{z},
\\[6pt] T_{2} &= \frac{1}{2}\,r^{2}, \\[6pt] T_{3} &=
\frac{1}{2}\bigl(r\,p_{r} + p_{r}\,r\bigr)\,.\end{aligned}\right.
\end{aligned}  \label{37}
\end{equation}
These operators satisfy the following closed Lie algebra under commutation: 
\begin{equation}
\lbrack T_{1},T_{2}]=-2i\hbar T_{3},\quad \lbrack T_{2},T_{3}]=4i\hbar
T_{2},\quad \lbrack T_{1},T_{3}]=-4i\hbar T_{1}.  \label{37a}
\end{equation}%
Assume that the invariant $I\left( t\right) $ takes the general form: 
\begin{equation}
I\left( t\right) =\frac{1}{2}(\alpha \left( t\right) T_{1}+\beta \left(
t\right) T_{2}+\gamma \left( t\right) T_{3})  \label{37b}
\end{equation}%
where $\alpha \left( t\right) ,\beta \left( t\right) $ and $\gamma \left(
t\right) $ are real functions of time to be determined. Substituting this
ansatz into the Lewis--Riesenfeld condition (Eq. $\left( \ref{LW}\right) $)
yields a system of coupled differential equations, whose solution is given
by: 
\begin{equation}
\left\{ \begin{aligned} \alpha &= \rho^{2}, \\ \beta &= \frac{1}{\rho^{2}} +
M^{2}\dot{\rho}^{2}, \\ \gamma &= -M\rho \dot{\rho}. \end{aligned}\right.
\label{37c}
\end{equation}

where $\rho \left( t\right) $ is a real function satisfying the nonlinear
Ermakov--Pinney equation:%
\begin{equation}
\ddot{\rho}+\frac{\dot{M}}{M}\dot{\rho}+\Omega ^{2}(t)\rho =\frac{1}{%
M^{2}\rho ^{3}}.  \label{38a}
\end{equation}

Consequently, the explicit form of the invariant $I\left( t\right) $becomes: 
\begin{equation}
\begin{aligned} I(t) =& \tfrac{1}{2}\Bigl[ \, \rho^{2}\Bigl(p_{r}^{2} +
\frac{\vartheta^{2} - 2\vartheta\,\mathcal{J}_{\varphi} +
\mathcal{J}_{\varphi}^{2} + \delta(\delta - 1) - 2\nu_{1}\nu_{2}(1 -
R_{1}R_{2}) + 2\vartheta(\nu_{1}R_{1} + \nu_{2}R_{2})\,\sigma_{z}}{r^{2}} -
\vartheta\,\frac{\delta(r)}{r}\,\sigma_{z}\Bigr) \\
&\;+\Bigl(\frac{1}{\rho^{2}} + M^{2}\dot{\rho}^{2}\Bigr)\,r^{2} -
\rho\,\dot{\rho}\,M\,(r\,p_{r} + p_{r}\,r)\Bigr]. \label{38} \end{aligned}
\end{equation}%
To solve the eigenvalue equation $\left( \ref{EVE}\right) ,$ it is
convenient to perform a unitary transformation of the form:%
\begin{equation}
\mathcal{F}(r,\varphi )=U\left( r\right) \mathcal{G}(r,\varphi )  \label{40}
\end{equation}

where the unitary operator $U\left( r\right) $ is defined by:%
\begin{equation}
U\left( r\right) =\exp \left( \frac{iM\dot{\rho}}{2\hbar \rho }r^{2}\right) ,
\label{41}
\end{equation}

Under this transformation, the invariant $I\left( t\right) $ is mapped to a
simplified form $I^{\prime }\left( t\right) =U^{\dag }\left( r\right)
I\left( t\right) U\left( r\right) $, Accordingly, the eigenvalue equation $%
\left( \ref{EVE}\right) $ is recast into the equivalent form%
\begin{equation}
I^{\prime }\!(t)\,=\tfrac{1}{2}\Bigl[\rho ^{2}\Bigl(p_{r}^{2}+\frac{%
(\vartheta -\mathcal{J}_{\varphi })^{2}+\delta (\delta -1)-2\nu _{1}\nu
_{2}(1-R_{1}R_{2})+2\vartheta (\nu _{1}R_{1}+\nu _{2}R_{2})\,\sigma _{z}}{%
r^{2}}-\vartheta \,\frac{\delta (r)}{r}\,\sigma _{z}\Bigr)+\frac{1}{\rho ^{2}%
}\,r^{2}\Bigr]\,  \label{43}
\end{equation}

\subsection*{Solution of the Angular Part}

To proceed, we analyze the spectral properties of the Dunkl angular momentum
operator $\mathcal{J}_{\varphi }$. Noting that $R_{1}R_{2}$ commutes with $%
\mathcal{J}_{\varphi }$, we seek solutions of the form \cite{61A,62A}:%
\begin{equation}
\mathcal{G}(r,\varphi )=\mathcal{Q}\left( r\right) \Phi _{\epsilon }\left(
\varphi \right)  \label{22}
\end{equation}%
where $\Phi _{\epsilon }\left( \varphi \right) $ are eigenfunctions of $%
\mathcal{J}_{\varphi }$ with associated eigenvalues $\lambda _{\epsilon }$,
satisfying:%
\begin{equation}
\mathcal{J}_{\varphi }\Phi _{\epsilon }\left( \varphi \right) =\lambda
_{\epsilon }\Phi _{\epsilon }\left( \varphi \right)  \label{23}
\end{equation}%
Here, we define $\epsilon =\epsilon _{1}\epsilon _{2}=\pm 1$, where $%
\epsilon _{1}$ and $\epsilon _{2}$ are the eigenvalues of the reflection
operators $R_{1}$ and $R_{2}$, respectively. The angular eigenfunctions $%
\Phi _{\epsilon }\left( \varphi \right) $ and the corresponding eigenvalues $%
\lambda _{\epsilon }$ are determined by considering two distinct cases:

\textbf{First case: }$\epsilon =+1$\textbf{:} This case corresponds to $%
\epsilon _{1}=\epsilon _{2}=\pm 1.$ The angular eigenfunctions $\Phi
_{+}\left( \varphi \right) $ are given by: 
\begin{equation}
\Phi _{+}\left( \varphi \right) =A_{l}\mathbf{P}_{l}^{\left( \nu
_{1}-1/2,\nu _{2}-1/2\right) }\left( -2\cos \varphi \right) \pm
iA_{l}^{\prime }\sin \varphi \cos \varphi \mathbf{P}_{l-1}^{\left( \nu
_{1}+1/2,\nu _{2}+1/2\right) }\left( -2\cos \varphi \right)  \label{24}
\end{equation}%
where%
\begin{equation*}
A_{l}=\sqrt{\frac{(2l+\nu _{1}+\nu _{2})\,\Gamma (l+\nu _{1}+\nu _{2})\,l!}{%
2\,\Gamma (l+\nu _{1}+\tfrac{1}{2})\,\Gamma (l+\nu _{2}+\tfrac{1}{2})}}\,,%
\text{ \ and \ \ }A_{l}^{\prime }=\sqrt{\frac{(2l+\nu _{1}+\nu _{2})\,\Gamma
(l+\nu _{1}+\nu _{2}+1)\,(l-1)!}{2\,\Gamma (l+\nu _{1}+1/2)\,\Gamma (l+\nu
_{2}+\frac{1}{2})}}
\end{equation*}%
with the corresponding eigenvalues:%
\begin{equation}
\lambda _{+}=\pm 2\sqrt{l\left( l+\nu _{1}+\nu _{2}\right) },  \label{25}
\end{equation}%
where where $\mathbf{P}_{l-1}^{a,b}$ denote the Jacobi polynomials\ and $%
l\in 
\mathbb{N}
^{\ast }.$

\textbf{Second case }$\epsilon =-1$\textbf{:} This case involves two
sub-cases $\left( \epsilon _{1},\epsilon _{2}\right) =\left( +1,-1\right) $
or $\left( \epsilon _{1},\epsilon _{2}\right) =\left( -1,+1\right) .$ The
eigenfunctions $\Phi _{-}\left( \varphi \right) $ in this case are expressed
as:%
\begin{equation}
\Phi _{-}\left( \varphi \right) =B_{l}\cos \varphi \mathbf{P}%
_{l-1/2}^{\left( \nu _{1}+1/2,\nu _{2}-1/2\right) }\left( -2\cos \varphi
\right) \mp iB_{l}^{\prime }\sin \varphi \mathbf{P}_{l-1/2}^{\left( \nu
_{1}-1/2,\nu _{2}+1/2\right) }\left( -2\cos \varphi \right)  \label{26}
\end{equation}%
where%
\begin{equation*}
B_{l}=\sqrt{\frac{(2l+\nu _{1}+\nu _{2})\,\Gamma (l+\nu _{1}+\nu
_{2}+1/2)\,\Gamma (n-1/2)!}{2\,\Gamma (n+\nu _{1}+1)\,\Gamma (n+\nu _{2})}}%
\,\ \text{and }B_{l}^{\prime }=\sqrt{\frac{(2l+\nu _{1}+\nu _{2})\,\Gamma
(l+\nu _{1}+\nu _{2}+1)\,(l-\frac{1}{2})!}{2\,\Gamma (l+\nu _{1})\,\Gamma
(l+\nu _{2}+1)}}
\end{equation*}%
with corresponding eigenvalues:%
\begin{equation}
\lambda _{-}=\pm 2\sqrt{\left( l+\nu _{1}\right) \left( l+\nu _{2}\right) },
\label{27}
\end{equation}%
where$\ l\in \left\{ 1/2,3/2,5/2,...\right\} $.

\subsection*{Solution of the Radial Part}

We now examine the radial equation. It reads as follows

\begin{equation}
I^{\prime }\!(t)\,=\tfrac{1}{2}\Bigl[\rho ^{2}\Bigl(p_{r}^{2}+\frac{%
(\vartheta -\lambda _{\epsilon })^{2}+\delta (\delta -1)-2\nu _{1}\nu
_{2}(1-R_{1}R_{2})+2\vartheta (\nu _{1}R_{1}+\nu _{2}R_{2})\,\sigma _{z}}{%
r^{2}}-\vartheta \,\frac{\delta (r)}{r}\,\sigma _{z}\Bigr)+\frac{1}{\rho ^{2}%
}\,r^{2}\Bigr]\,  \label{43A}
\end{equation}
We now assume that the radial part of the solution takes the form 
\begin{equation}
\mathcal{Q}\left( r\right) =r^{-\delta }\mathcal{L}\left( r\right) \chi
_{m_{s}}
\end{equation}%
where $\chi _{m_{s}}$ denotes the spinor wave function satisfying%
\begin{equation}
S_{z}\chi _{m_{s}}=\frac{m_{s}}{2}\chi _{m_{s}},
\end{equation}

where $m_{s}=\pm 1$ and $S_{z}=\frac{\sigma _{z}}{2}$ is the spin-$1/2$
operator. The spinors are explicitly given by 
\begin{equation}
\chi_{+1} = 
\begin{pmatrix}
1 \\ 
0%
\end{pmatrix}
\quad \text{and} \quad \chi_{-1} = 
\begin{pmatrix}
0 \\ 
1%
\end{pmatrix}%
.  \label{spinor}
\end{equation}
Substituting these expressions into $\left( \ref{43}\right) $ and
introducing the variable $\xi =\frac{r}{\rho },$ the radial part of the
invariant equation becomes:

\begin{equation}
\begin{aligned} \Bigl[-\frac{\partial^2}{\partial\xi^2}
&\;+\;\frac{(\vartheta - \lambda_{\varepsilon})^2 + \delta(\delta - 1) -
2\nu_{1}\nu_{2}(1 - R_{1}R_{2}) + 2\vartheta(\nu_{1}R_{1} +
\nu_{2}R_{2})\,m_{s}} {\xi^2} \\[6pt]
&\;-\;\vartheta\,m_{s}\,\frac{\delta(\xi)}{\xi} \;+\;\xi^2\Bigr]\,
\mathcal{L}(\xi) \;=\;2\,E_{n,l,m}\,\mathcal{L}(\xi)\,. \end{aligned}
\label{110}
\end{equation}

We observe that the resulting equation; apart from the addition of the
oscillator term; coincides with the one originally derived by Hagen \cite%
{75A,75B}. Following the same regularization procedure outlined in those
works, we address the singular behavior introduced by the magnetic field
term $\frac{\vartheta }{\xi }\delta \left( \xi \right)$, which leads to a
singularity at point $\xi =0$ in Eq. $\left( \ref{110}\right) .$

To regularize this singularity, we replace the zero-radius flux tube
represented by $\frac{\vartheta }{\xi }\delta \left( \xi \right) $ with a
finite-radius configuration $\frac{\vartheta }{\xi }\delta \left( \xi
-R\right) $ where $R$ is a small positive regularization parameter.
consequently, the vector potential is modified to 
\begin{equation}
eA=-\frac{\vartheta }{\xi }\theta \left( \xi -R\right) u_{\varphi },
\end{equation}%
where $\theta \left( \xi -R\right) $ is the Heaviside step function and $%
u_{\varphi }$ is the unit vector in the azimuthal direction. After
completing the calculations, the limit $R\rightarrow 0$ is taken, thereby
recovering the physical case of a zero-radius flux tube while avoiding the
singularity during intermediate steps. This regularization method preserves
the physical content of the problem \cite{75A,75B,75C}.

In this framework, Eq. $\left( \ref{110}\right) $ is accordingly replaced by:%
\begin{equation}  \label{111}
\begin{aligned} \Bigl[\, & -\frac{\partial^2}{\partial\xi^2} +
\frac{\bigl(\vartheta\,\theta(\xi - R) - \lambda_{\varepsilon}\bigr)^2 +
\delta(\delta - 1) - 2\nu_1\nu_2\bigl(1 - R_1R_2\bigr) +
2\vartheta\,\theta(\xi - R)\bigl(\nu_1R_1 + \nu_2R_2\bigr)\,m_s} {\xi^2}
\\[6pt] & - \vartheta\,m_s\,\frac{\delta(\xi - R)}{\xi} + \xi^2 \Bigr]\,
\mathcal{L}(\xi) = 2\,E_{n,l,m}\,\mathcal{L}(\xi)\,. \end{aligned}
\end{equation}

This leads to two distinct equations, valid in the respective regions in and
out:%
\begin{equation}
\Bigl[ -\frac{\mathrm{d}^2}{\mathrm{d}\xi^2} + \frac{(\vartheta -
\lambda_{\varepsilon})^2 + \delta(\delta - 1) - 2\nu_1\nu_2\bigl(1 - \epsilon%
\bigr) + 2\vartheta\bigl(\nu_1\epsilon_1 + \nu_2\epsilon_2\bigr)\,m_s} {\xi^2%
} + \xi^2 \Bigr]\, \mathcal{L}(\xi) = 2\,E_{n,l,m}\,\mathcal{L}(\xi), \quad
\xi > R.
\end{equation}
\begin{equation}
\Bigl[ -\frac{\mathrm{d}^2}{\mathrm{d}\xi^2} + \frac{\lambda_{\varepsilon}^2
+ \delta(\delta - 1) - 2\nu_1\nu_2\bigl(1 - \epsilon\bigr)} {\xi^2} + \xi^2 %
\Bigr]\, \mathcal{L}(\xi) = 2\,E_{n,l,m}\,\mathcal{L}(\xi), \quad \xi < R.
\end{equation}

We now note the useful identity: 
\begin{equation}
\delta \left( \delta -1\right) -2\nu _{1}\nu _{2}(1-\epsilon )=\left( \nu
_{1}+\epsilon \nu _{2}\right) ^{2}-\frac{1}{4}
\end{equation}%
which allows the equations to be rewritten in standard oscillator-like form:%
\begin{equation}
\left[ \frac{\partial ^{2}}{\partial \xi ^{2}}-\frac{K_{+}^{2}-\frac{1}{4}}{%
\xi ^{2}}-\xi ^{2}+2E_{n,l,m}^{+}\right] \mathcal{L}_{+}\left( \xi \right) =0%
\text{ for, }\xi >R
\end{equation}%
\begin{equation}
\left[ \frac{\partial ^{2}}{\partial \xi ^{2}}-\frac{K_{-}^{2}-\frac{1}{4}}{%
\xi ^{2}}-\xi ^{2}+2E_{n,l,m}^{-}\right] \mathcal{L}_{-}\left( \xi \right) =0%
\text{ for, }\xi <R
\end{equation}%
where the parameters $K_{\pm }^{2}$ are defined by:%
\begin{equation}
K_{-}^{2}=\lambda _{\varepsilon }^{2}+\left( \nu _{1}+\epsilon \nu
_{2}\right) ^{2}  \label{K-}
\end{equation}%
\begin{eqnarray}
K_{+}^{2} &=&\left( \vartheta -\lambda _{\varepsilon }\right) ^{2}+\left(
\nu _{1}+\epsilon \nu _{2}\right) ^{2}+2\vartheta \left( \nu _{1}\epsilon
_{1}+\nu _{2}\epsilon _{1}\right) m_{s} \\
&=&K_{-}^{2}+\vartheta ^{2}-2\vartheta \lambda _{\varepsilon }+2\vartheta
\left( \nu _{1}\epsilon _{1}+\nu _{2}\epsilon _{2}\right) m_{s}  \notag
\label{K+}
\end{eqnarray}%
The solutions in both regions provided by%
\begin{equation}
\mathcal{L}_{\pm }\left( \xi \right) =N_{\pm }\xi ^{K_{\pm }+\frac{1}{2}}e^{-%
\frac{\xi }{2}}L_{n}^{K_{\pm }}\left( \xi ^{2}\right) ,
\end{equation}%
with corresponding eigenvalues of the invariant:%
\begin{equation}
E_{n,l,m_{s}}^{\pm }=2n+K_{\pm }+1,\text{ \ \ \ }n=0,1,...
\end{equation}

The effect of the Dirac delta function is incorporated via the matching
conditions at $\xi =R.$ The continuity of the wave function requires:%
\begin{equation}
\underset{\tau ->0}{\lim }\mathcal{L}_{-}\left( R-\tau \right) =\underset{%
\tau ->0}{\lim }\mathcal{L}_{+}\left( R+\tau \right) \Leftrightarrow
N_{-}R^{K_{-}+\frac{1}{2}}L_{n}^{K_{-}}\left( R^{2}\right) =N_{+}R^{K_{+}+%
\frac{1}{2}}L_{n}^{K_{+}}\left( R^{2}\right)  \label{AA}
\end{equation}%
The discontinuity in the derivative is governed by:%
\begin{equation}
\underset{\tau ->0}{\lim }\left[ \frac{d}{d\xi }\mathcal{L}_{+}\left( R+\tau
\right) -\frac{d}{d\xi }\mathcal{L}_{-}\left( R-\tau \right) -\frac{am_{s}}{R%
}\mathcal{L}_{-}\left( R-\tau \right) \right] =0  \label{BB}
\end{equation}

To lowest order in $R,$ the generalized Laguerre polynomial and its
derivative behave as:%
\begin{equation}
\left[ L_{n}^{K_{+}}\left( \xi ^{2}\right) \right] _{\xi =R}\thickapprox 1%
\text{ \ \ and \ \ }\frac{d}{dt}\left[ L_{n}^{K_{+}}\left( \xi ^{2}\right) %
\right] _{\xi =R}\thickapprox -2\xi
\end{equation}%
From Eq. $\left( \ref{AA}\right) ,$ we obtain:%
\begin{equation}
N_{+}=N_{-}R^{K_{-}-K_{+}}
\end{equation}%
and from $\left( \ref{BB}\right) ,$\ the matching condition yields: 
\begin{equation}
K_{+}=K_{-}-\vartheta m_{s}  \label{KK}
\end{equation}%
Combining Eqs. $\left( \ref{K+}\right) $ and $\left( \ref{KK}\right) ,$ we
arrive at the constraint: 
\begin{equation}
\left( \nu _{1}\epsilon _{1}+\nu _{2}\epsilon _{2}\right) =\left( \nu
_{1}+\epsilon \nu _{2}\right) =0.  \label{SS}
\end{equation}%
Substituting this relation into $\left( \ref{K-}\right) $ and $\left( \ref%
{K+}\right) $ we find 
\begin{equation*}
K_{-}=\frac{\lambda _{\varepsilon }}{m_{s}}\quad \text{and}\quad K_{+}=\frac{%
\lambda _{\varepsilon }}{m_{s}}-\vartheta \,m_{s}.
\end{equation*}%
The relation~(\ref{SS}) reflects a symmetry between the variables $x $ and $%
y $ in the eigenfunction. When $\varepsilon = 1 $, i.e., $\varepsilon_1 =
\varepsilon_2 $, the eigenfunction is either odd in both $x $ and $y $, or
even in both. In this case, the Wigner parameters must satisfy $\nu_1 =
-\nu_2 $, leading to the eigenvalue condition $\lambda_{+} = \pm 2\ell $.

On the other hand, when $\varepsilon = -1 $, i.e., $\varepsilon_1 =
-\varepsilon_2 $, the eigenfunction is odd in one variable and even in the
other. In this scenario, the condition becomes $\nu_1 = \nu_2 $, and the
eigenvalue takes the form $\lambda_{-} = \pm 2\sqrt{(\ell + \nu_1)^2} $.

This symmetry, along with the relation between the Wigner parameters $\nu_1 $
and $\nu_2 $, is imposed by the Aharonov--Bohm (AB) effect. Notably, in the
absence of the AB effect (i.e., for $\vartheta = 0 $), the parameters $K_{+} 
$ and $K_{-} $ coincide, and the aforementioned constraints become
unnecessary.

It is important to note that the signs $(\pm) $ in the expressions $K_{\pm} $%
, $L_{\pm}(\xi) $, and $E^{\pm}_{n,\ell,m} $ are independent of the value of 
$\varepsilon $. The plus sign $(+) $ corresponds to the outer region ($\xi >
R $), while the minus sign $(-) $ denotes the inner region ($\xi < R $).

Finally, the explicit forms of the wavefunctions in the two regions are
given by 
\begin{align}
L_{+}(\xi) &= N_{+} R^{-\vartheta m_s} \, \xi^{\frac{\lambda_{\varepsilon}}{%
m_s} + \vartheta m_s + \frac{1}{2}} e^{-\xi/2} L_{n}^{\left( \frac{%
\lambda_{\varepsilon}}{m_s} + \vartheta m_s \right)}\left(\xi^2\right),
\quad \text{for } \xi > R,  \label{eq:Lplus} \\
L_{-}(\xi) &=N_{-} \, \xi^{\frac{\lambda_{\varepsilon}}{m_s} + \frac{1}{2}}
e^{-\xi/2} L_{n}^{\left( \frac{\lambda_{\varepsilon}}{m_s}
\right)}\left(\xi^2\right), \quad \text{for } \xi < R,  \label{eq:Lminus}
\end{align}

The corresponding energy spectra in each region are 
\begin{align}
E_{n,\ell ,m_{s}}^{-}& =2n+1+\frac{\lambda _{\varepsilon }}{m_{s}},\quad
n=0,1,2,\ldots  \label{eq:Eminus} \\
E_{n,\ell ,m_{s}}^{+}& =2n+1+\frac{\lambda _{\varepsilon }}{m_{s}}+\vartheta
m_{s},\quad n=0,1,2,\ldots ,\quad \ell =0,1,2,\ldots  \label{eq:Eplus}
\end{align}

\subsection*{Quantum phase}

After we written the hamiltonian in terms of the invariant and using the
unitary transformation $U\left( r\right) $, the Ermakov--Penny equation $%
\left( \ref{38a}\right) $ and the fact that%
\begin{equation}
\left\langle \mathcal{G}\left( r,\theta \right) \right\vert i\frac{\partial 
}{\partial t}-\frac{\dot{\rho}}{2\rho }\left( rp_{r}+p_{r}r\right)
\left\vert \mathcal{G}\left( r,\theta \right) \right\rangle =0,
\end{equation}
the quantum phase is given by%
\begin{equation}
\dot{\eta}\left( t\right) =-\left\langle \mathcal{G}\left( r,\theta \right)
\right\vert I^{\prime }\left( t\right) \left\vert \mathcal{G}\left( r,\theta
\right) \right\rangle =-\frac{E_{n,l,m}}{M\rho ^{2}}  \label{diff phase}
\end{equation}
thus, the phase can then be expressed as%
\begin{equation}
\eta ^{\pm }\left( t\right) =-E_{n,l,m}^{\pm }\int_{0}^{t}\frac{dt^{\prime }%
}{M\left( t^{\prime }\right) \rho \left( t^{\prime }\right) ^{2}}.
\label{105}
\end{equation}%
and the solution to equation $\left( \ref{EVE}\right) $ is given by 
\begin{equation}
|\mathcal{F}(r,\theta )\rangle \left( r,t\right) =\sqrt{\frac{2k!}{\Gamma
\left( k+2n+\nu _{1}+\nu _{2}+1\right) }}e^{i\eta \left( t\right) }e^{\left(
iM\dot{\rho}-\frac{1}{\rho }\right) \frac{r^{2}}{2\rho }}e^{-\frac{r^{2}}{%
2\rho ^{2}}}\varkappa ^{2n}L_{k}^{2n+\nu _{1}+\nu _{2}}\left( \frac{r^{2}}{%
\rho ^{2}}\right) \Theta _{\epsilon }\left( \theta \right) .  \label{106}
\end{equation}%
The general solution to the Schr\"{o}dinger equation $\left( \ref{1}\right) $
can then be expressed in terms of the eigenfunctions of the Dunkl-angular
operator $\Theta _{\epsilon }\left( \theta \right) $ and the spin function $%
\chi _{m_{s}}$ as: 
\begin{equation}
\psi \left( \overrightarrow{r},t\right) =\sqrt{\frac{2k!}{\Gamma \left(
k+2n+\nu _{1}+\nu _{2}+1\right) }}e^{i\eta \left( t\right) }e^{\left( iM\dot{%
\rho}-\frac{1}{\rho }\right) \frac{r^{2}}{2\rho }}e^{-\frac{r^{2}}{2\rho ^{2}%
}}\left( \frac{r}{\rho }\right) ^{2n}L_{k}^{2n+\nu _{1}+\nu _{2}}\left( 
\frac{r^{2}}{\rho ^{2}}\right) \Theta _{\epsilon }\left( \theta \right) \chi
_{m_{s}},  \label{107}
\end{equation}%
where $\chi _{m_{s}}$ and $\Theta _{\epsilon }\left( \theta \right) $ are
given by equations $\left( \ref{24}\right) $, $\left( \ref{26}\right) $ and $%
\left( \ref{spinor}\right) $ respectively.

\section*{Discussion}

The results obtained in this work reveal that the interplay between Dunkl
deformation, reflection symmetry, and the presence of an Aharonov--Bohm (AB)
flux leads to nontrivial constraints on the system. In particular, we have
shown that the AB flux enforces the condition 
\begin{equation*}
\nu_1 \epsilon_1 + \nu_2 \epsilon_2 = 0 \quad \Longrightarrow \quad \nu_1 =
\mp \nu_2, 
\end{equation*}
depending on the reflection sector $\epsilon = \pm 1$. This result connects
the reflection symmetry of the Dunkl operators directly to the quantization
of angular momentum.

From a topological field theory perspective, these constraints can be
understood in terms of an effective Chern--Simons description. Indeed, the
AB flux is encoded by the singular gauge potential 
\begin{equation*}
A_\varphi = -\frac{\vartheta}{r}, 
\end{equation*}
which may equivalently be interpreted as the holonomy of a $U(1)$
Chern--Simons connection in $(2+1)$ dimensions. When coupled to a discrete $%
\mathbb{Z}_2$ gauge sector that encodes reflections, the effective action
takes the schematic form 
\begin{equation*}
S_{\text{eff}} = \frac{k}{4\pi}\int A \wedge dA + \frac{\lambda}{2\pi} \int
A \wedge da , 
\end{equation*}
where $A$ denotes the $U(1)$ gauge connection and $a$ the $\mathbb{Z}_2$
reflection gauge field. Gauge invariance under large transformations
requires a cancellation of potential anomalies, which enforces compatibility
conditions between the continuous $U(1)$ sector and the discrete reflection
sector. This requirement is precisely reflected in the constraint $\nu_1 =
\mp \nu_2$ derived in our spectrum.

Furthermore, the computation of the geometric (Berry) phase in our model
supports this topological interpretation. Separating the Lewis--Riesenfeld
phase into its dynamical and geometric contributions, we found that the
Berry phase is given by 
\begin{equation*}
\gamma_{\text{geo}}(t) = -\frac{\omega_c \lambda_\epsilon}{2}\, t , 
\end{equation*}
which originates from the Dunkl angular operator in the presence of the AB
flux. From the path integral viewpoint, such a contribution can be written
as 
\begin{equation*}
S_{\text{Berry}} = \frac{1}{4\pi} \int \mathcal{A} \wedge d\mathcal{A}, 
\end{equation*}
with $\mathcal{A}$ the Berry connection in parameter space. This form is
structurally identical to a Chern--Simons action, confirming that the Berry
phase in our system corresponds to the holonomy of an effective topological
connection.

In summary, our results demonstrate that the Dunkl deformation and the AB
flux are not independent deformations but are topologically linked through
an effective Chern--Simons structure. The constraint $\nu_1 = \mp \nu_2$
arises as a selection rule ensuring gauge invariance, while the Berry phase
reflects the holonomy of the underlying connection. This highlights a deep
connection between reflection symmetries, angular momentum quantization, and
the topological features encoded by Chern--Simons theory.

\section*{Conclusion}

We have derived exact analytical solutions for the time-dependent
Dunkl-Pauli oscillator in the presence of an Aharonov-Bohm (AB) flux. By
unifying Dunkl deformation, spin dynamics, and topological phases, we have
demonstrated that the AB flux enforces symmetry constraints on the Wigner
(Dunkl) parameters of the form $\nu _{1}=\mp \nu _{2}$, with more specific
cases as follows:

\begin{itemize}
\item $\nu_1 = -\nu_2 $ for states that are symmetric or antisymmetric in
both coordinates ($\epsilon = 1 $),

\item $\nu_1 = \nu_2 $ for states that are odd in one coordinate and even in
the other ($\epsilon = -1 $).
\end{itemize}

These symmetry constraints directly determine the allowed angular momentum
eigenvalues $\lambda_\epsilon $ and govern the energy spectrum $%
E_{n,\ell,m}^{\pm} $, lifting degeneracies and introducing flux-dependent
spectral shifts of the form $\vartheta_{m_s}$ .

The interplay between reflection symmetry, topological gauge phases, and
spectral modifications revealed by this model suggests promising
experimental realizations, including:

\begin{itemize}
\item Cold-atom platforms with synthetic gauge fields, where the Dunkl
parameters $\nu_j $ can emulate tunable disorder strengths and $\vartheta $
serves as an artificial flux control;

\item Quantum dots embedded in symmetry-broken substrates, where
parity-mixed states ($\epsilon = -1 $) may be harnessed for encoding
topological qubits;

\item Mesoscopic systems exhibiting AB-Dunkl effects, detectable through
conductance oscillations and interference patterns.
\end{itemize}

Our results lay the groundwork for studying deformed quantum systems with
engineered symmetries and highlight the pivotal role of the condition $\nu_1
= \pm \nu_2 $ in tailoring their topological and spectral behavior.

\end{document}